\documentclass[a4paper]{article}

\usepackage{INTERSPEECH2022}
\usepackage[para,online,flushleft]{threeparttable}
\usepackage{multirow}
\usepackage{subcaption}
\usepackage{graphicx}
\usepackage{adjustbox}
\usepackage{hyperref}

\title{Deep Neural Convolutive Matrix Factorization for Articulatory Representation Decomposition}
\name{Jiachen Lian$^{1}$, Alan W Black$^{2}$,  Louis Goldstein$^{3}$, Gopala Krishna Anumanchipalli$^{1}$}
  
\address{$^{1}$ UC Berkeley, CA $^{2}$ Carnegie Mellon University, PA $^{3}$ University of Southern California, CA}
\email{jiachenlian@berkeley.edu, awb@cs.cmu.edu, louisgol@usc.edu, gopala@berkeley.edu,}

\begin{document}
\maketitle
\begin{abstract}
 Most of the research on data-driven speech representation learning has focused on raw audios in an end-to-end manner, paying little attention to their internal phonological or gestural structure. This work, investigating the speech representations derived from articulatory kinematics signals, uses a neural implementation of convolutive sparse matrix factorization to decompose the articulatory data into interpretable gestures and gestural scores. By applying sparse constraints, the gestural scores leverage the discrete combinatorial properties of phonological gestures. Phoneme recognition experiments were additionally performed to show that gestural scores indeed code phonological information successfully. The proposed work thus makes a bridge between articulatory phonology and deep neural networks to leverage informative, intelligible,  interpretable,and efficient speech representations. The code is made publicly available at \url{https://github.com/Berkeley-Speech-Group/ema_gesture}. 
\end{abstract}
\section{Introduction}



Research on speech representation learning has been dominated by deep learning recently in the areas such as speaker recognition~\cite{lian2020masked, ECAPA-TDNN}, automatic speech recognition~\cite{conformer}, voice conversion~\cite{DSVAE-VC, C-DSVAE} and text-to-speech~\cite{VITS, lian2022utts}, etc. The goal of speech representation learning is to optimize  both the performance of the model architectures and the interpretability of the learned representations. As there is growing demand of real-life applications of speech interfaces \cite{herff2016automatic}, the performance is emphasized to a larger extent, enabling  human-machine interactions highly accurate and robust. Consequently, in most of these works the interpretability of representations has not been explored to an equivalent extent, which is one of the most significant bottlenecks that keeps the speech research from going farther. In general, speech representations need to be better understood and developed. 

People usually represent speech via audio because human perceive speech through hearing and audio is cheap to record, collect and process. However, speech processing is quite a lot different from audio processing. It might not need any evidence to indicate that any information that can be perceived via human can be perceived anywhere from source to destination. Perceiving the speech signal from the source and leveraging how it is produced are the most straightforward way to interpret it. The speech signal is the result of respiratory, phonatory and articulatory processes that generate the perceivable acoustic resonances to encode an intended linguistic message \cite{origin-of-speech}. In that sense, perceiving the speech signal from articulatory data is a preferred way to derive interpretable, natural and robust speech representations. 

\begin{figure*}[!ht]
  \includegraphics[width=17cm,height=8cm]{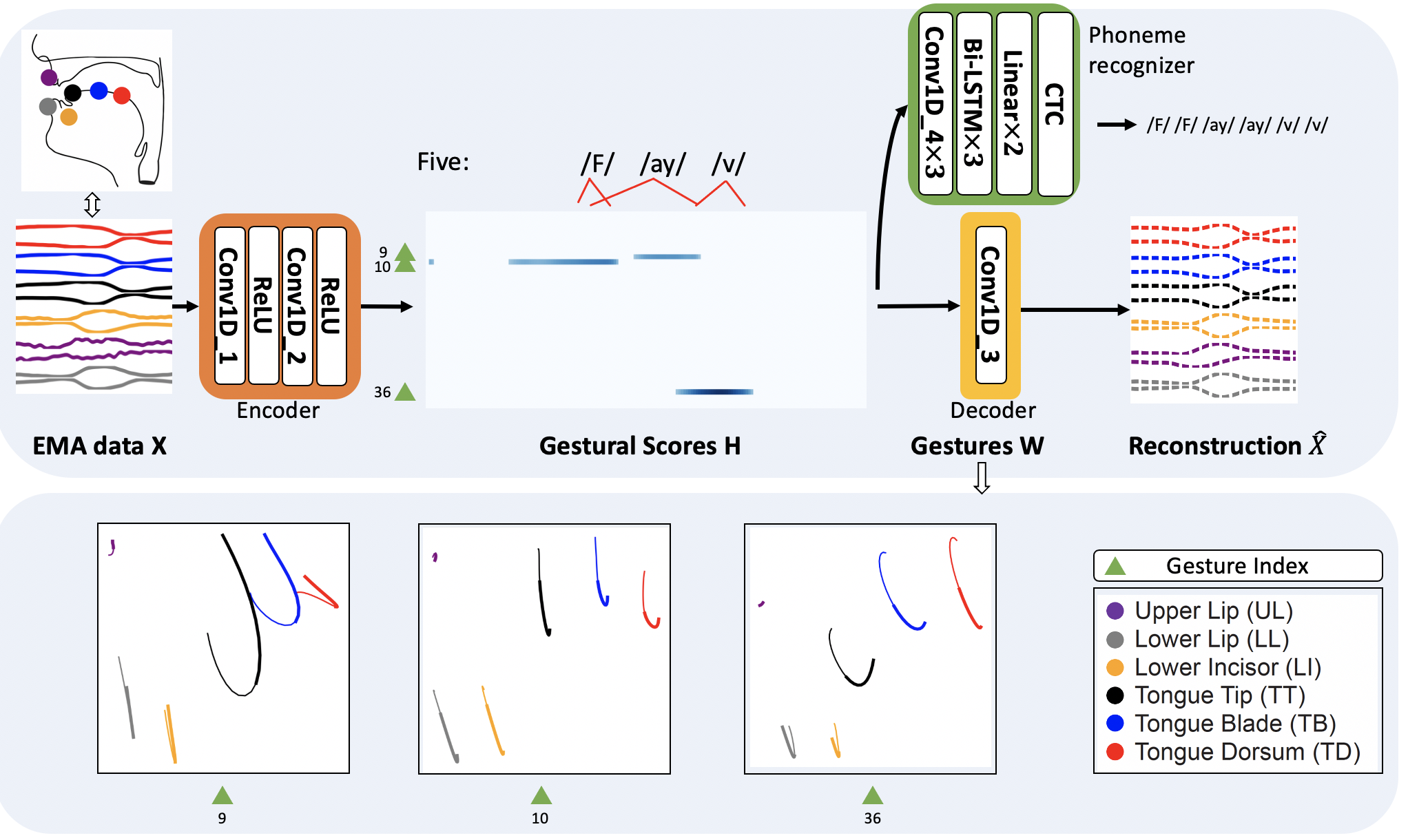}
  \caption{Neural Convolutive Matrix Factorization to Interpret Gestures and Gestural Scores.
  The panel in the bottom visualizes four activated gestures in the example utterance for \textit{"Five"} . The gestures capture the moving patterns of articulators. Each moving pattern goes from thinner line to thicker line capturing about 200ms.}
  \label{model}
\end{figure*}

The framework of articulatory phonology \cite{articulatory-phonology} has offered a lawful approach to modeling the relation between phonological representations as a set of discrete compositional units, or \textit{gestures}, and the variability in time that derives from variation in the activation of the gestures in real-time: the magnitude of their activation, and the temporal intervals of activation as represented in \textit{gestural scores}. However, the gestures and gestural scores of particular utterances have never been estimated in a completely data-driven manner. \cite{spatio-temporal} utilized the convolutive sparse non-negative matrix factorization (CSNMF) to decompose the non-negative articulatory data into the gestures and gestural scores, both of which are pretty much interpretable. The downsides of such method are that all the training utterances have to be concatenated into a large matrix, resulting in both memory and training efficiency issues. Additionally, such a model is not compatible with the modern deep learning based speech models so that it is challenging to perform end-to-end training on articulatory data. 

To handle the aforementioned problem, \cite{neural-nmf} proposed an auto-encoder based model to replace non-negatve matrix factorization for speech separation task. Inspired by this work, we propose a convolutional auto-encoder as the neural implementation of convolutive matrix factorization. Such auto-encoder based matrix factorization method is compatible with modern deep neural network and the batch-wise optimization improves the convergence rate to the huge extent. Under such framework, the articulatory signal is decomposed into \textit{gestures} and \textit{gestural scores} which are still interpretable. The gestural scores are the learned articulatory speech representations and are constrained to be sparse. In the last stage, the phoneme recognition experiments were performed to show that the learned gestural scores are also intelligible and consistent in time domain. All the experiments are performed using MNGU0 EMA (Electromagnetic midsagittal articulography) \cite{mngu0} corpus. The intention is that the proposed work could bridge the gap between explainable articulatory phonology and modern deep neural networks to deliver interpretable, intelligible, informative, and efficient speech representations.
\section{Proposed Methods}
\subsection{Neural Convolutive Sparse Matrix Factorization}
Denote EMA data as $X\in \mathbb{R}^{C \times t}$, where (C,t) is (number of channels, segment length). By convolutive matrix factorization \cite{spatio-temporal}:
\begin{equation} \label{CSNMF}
    X\approx \Sigma_{i=0}^{T-1}W(i)\cdot \overrightarrow{H}^i
\end{equation}
$W\in \mathbb{R}^{T\times C \times D}$ is gestures and $H\in \mathbb{R}^{D\times t}$ is gestural scores, where $D$ is number of gestures and $T$ is the kernel size. $\overrightarrow{H}^i$ indicates that $i$ columns of $H$ are shifted to the right. 

It is observable that Eq. \ref{CSNMF} is actually the 1-d convolution with kernel $W$ and input matrix $H$. By auto-encoder matrix factorization \cite{neural-nmf}, H should be the hidden representation derived from the encoder which takes the pseudo-inverse of $W$ as parameters. However, calculating the pseudo-inverse of high dimensional matrix is challenging. We experimentally justified that the encoder can be any types of neural networks with any number of layers. The proposed neural convolutive sparse matrix factorization is formularized as follows:
\begin{equation}
    H=max(f(X),0)
\end{equation}
\begin{equation}
    \hat{X}=W\odot H
\end{equation}
where $f(.)$ denotes any type of neural network. In the original non-negative matrix factorization problem, all components ($X,W,H$) have to be non-negative. However, in such neural implementation, only $H$ is required to be non-negative so that the gestures are always additive. There is no constraint for $W$ and $X$. 

\subsection{Loss Objectives}
\begin{figure}[!ht]
  \begin{subfigure}[b]{0.15\textwidth}
    \includegraphics[width=\textwidth]{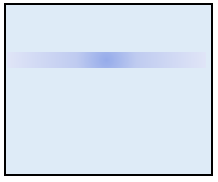}
    \caption{Low Entro}
  \end{subfigure}
  \begin{subfigure}[b]{0.15\textwidth}
    \includegraphics[width=\textwidth]{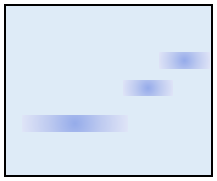}
    \caption{Medium Entro}
  \end{subfigure}
   \begin{subfigure}[b]{0.15\textwidth}
    \includegraphics[width=\textwidth]{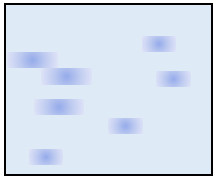}
    \caption{High Entro}
  \end{subfigure}
  \caption{Three types of gestural score matrices with the same sparsity but different entropy values.}
  \label{entropy of H}
\end{figure}

There are a couple of items in the loss function. The first one is the reconstruction loss, which is L2 loss. The second one is sparseness. According to \cite{sparseness}, the sparseness of a vector in time dimension and channel dimension are defined as in Eq.~(\ref{sparse time}) and Eq.~\ref{sparse channel} respectively:
\begin{equation} \label{sparse time}
S_1(H_i)= \frac{\sqrt{t}-\frac{L_1(H_i)}{L_2(H_i)}}{\sqrt{t}-1}  
\end{equation}
\begin{equation} \label{sparse channel}
S_2(H_i)= \frac{\sqrt{D}-\frac{L_1(H^T_i)}{L_2(H^T_i)}}{\sqrt{D}-1}  
\end{equation}
 where $H_i$and  $H^T_i$ denote the i-th row vector and column vector respectively. $L_1$ and $L_2$ denote $L_1$ norm and $L_2$ norm respectively. $n$ is the length of the vector. The sparseness of gestural score matrix in time dimension and channel dimension are shown in Eq.~(\ref{sparsity loss time dimension})(\ref{sparsity loss channel dimension}) respectively. The intuition is that at each time step, there are not too many gestures activated, and each gesture is not activate for a long time. 
\begin{equation}\label{sparsity loss time dimension}
S_1(H)=\frac{1}{D}\Sigma_{i=1}^{D}S(H_i)
\end{equation}
\begin{equation}\label{sparsity loss channel dimension}
S_2(H)=\frac{1}{D}\Sigma_{i=1}^{t}S(H^T_i)
\end{equation}
Following~\cite{wav2vec2}, we also introduce the entropy of the sparseness, denoted as:
 \begin{equation}
 E(H)=\frac{1}{D}\Sigma_{i=1}^{D}(-\frac{S_1(H_i)}{\Sigma_{i=1}^{D}S_1(H_i)}\log(\frac{S_1(H_i)}{\Sigma_{i=1}^{D}S_1(H_i)}))     
 \end{equation}
It should be noticed that the sparseness cannot control the number of gestures that are activated accurately. For instance, the H matrix with only one gesture activated for a long time interval might have the same sparsity with the matrix with multiple gestures activated for shorter time intervals. Typically we expect that a proper number of gestures should be activated. Fig. \ref{entropy of H} gives an intuition of entropy loss. All three H matrices have the same sparsity. If the entropy is pretty low, only one gesture is activated, which leaves many other gestures unused. If the entropy is  pretty high, some of activated gestures are redundant, which makes the gestural score less explainable. We do not consider the entropy loss on the channel dimension since it empirically does not make significant difference to the gestural score. 

We introduce balanced factors $\lambda_1, \lambda_2$ and $\lambda_3$ to limit both sparsity and entropy to a certain range. L2 loss is used as the reconstruction loss, as shown below:
\begin{equation} \label{rec loss}
    L_{rec}=||X-\hat{X}||_2
\end{equation}
For EMA resynthesis task, the loss function is shown in Eq. \ref{loss resynthesis}, where $\mathbb{E}_{X}$ means the loss is computed by taking the average in the mini-batch.
\begin{equation} \label{loss resynthesis}
    L_{res}=\mathbb{E}_{X}[L_{rec}-\lambda_1S_1(H)-\lambda_2S_2(H)+\lambda_3E(H)]
\end{equation}
For phoneme recognition experiments, CTC \cite{ctc} loss $L_{CTC} $ is used. For joint resynthesis-phoneme recognition task,
the loss function is shown as in Eq. \ref{loss joint}, where $\lambda_4$ is a balanced factor.
\begin{equation} \label{loss joint}
    L_{joint}=L_{res}+\lambda_4L_{CTC}
\end{equation}

\section{Experiments}
\subsection{Dataset}
MNGU0 EMA (Electromagnetic midsagittal articulography) \cite{mngu0} dataset is used in this work. There are in total 1263 utterances recorded from one single speaker. During the recording, six transducer coils were placed in the midsagittal plane at the upper lip, lower lip, lower incisors, tongue tip, tongue blade and tongue dorsum to record the coordinates (x and y) of their positions, and thus each EMA data frame takes 12 coordinates, as shown in Fig. \ref{model}. The sampling rate of EMA is 200 Hz. The Mel-Spectrogram is used as acoustic feature with the framing configuration of 25ms/16ms and feature dimension of 80. The unaligned phonemes extracted from text transcriptions via the CMU pronouncing dictionary\footnote{http://www.speech.cs.cmu.edu/cgi-bin/cmudict}, are used as labels for phoneme recognition task. 
The train/test split is 8:2, which is the same for all experiments. 

\subsection{Tasks and Evaluation Methods}
We perform two sets of experiments: (i) EMA Resynthesis. By resynthesizing the EMA data, we extract, visualize and interpret the gestures and gestural scores. The reconstruction loss $L_{rec}$ shown in Eq.~\ref{rec loss} averaged over all test samples is used to measure the \textit{informativeness} of gestural scores \cite{informationbottleneck}. The sparsity defined in Eq. \ref{sparsity loss time dimension} and \ref{sparsity loss channel dimension} is used to measure the \textit{efficiency} of gestural scores.
(ii) Phoneme Recognition (PR). PER (Phoneme Error Rate) is used as metric for this task. PR on EMA is performed to measure the \textit{intelligibility} of EMA data. PER on melspectrogram is performed to measure the \textit{intelligibility gap} between articulartory and acoustics data. Lastly, the joint training of EMA resynthesis and phoneme recognition on gestural scores is performed to measure the \textit{intelligibility} \cite{lakhotia2021generative} of learned sparse speech representations. Considering that EMA is not able to capture the difference between voiced and voiceless phones, we also relabel the phoneme sequence by assigning the same label to the phonemes with the same articulatory representation in EMA\footnote{Specifically,these tuples are expected to have the same articulatory labels: (p,b,m), (t,d,n), (ch,jh), (f,v), (sh,zh), (k,g,ng), (s,z), (th,dh)}, and compute PER on new labels. We call the latter metric as PER-V, which is reported for all PR experiments. The \textit{interpretability} of gestures and gestural scores is evaluated by subjective analysis given a set of utterances. We also subjectively measure the \textit{consistency} of the gestural scores via visualizing a set of phonetic units across different utterances. 

\subsection{Model Architectures}
The overall model backbone is shown in Fig. \ref{model}. The encoder takes EMA data $X$ in and outputs the gestural scores $H$. The decoder takes $H$ in and resynthesizes EMA data $\hat{X}$. For phoneme recognition or joint resynthesis-CTC experiments, the phoneme recognizer takes EMA, melspectrogram or $H$ in and predicts the alignment. Beamsearch algorithm is used for decoding with beam width of 50 in phoneme recognition task. 

\begin{table}[h]
    \centering

    \begin{adjustbox}{width=160pt,center}
    \begin{threeparttable}
    \begin{tabular}{ |ccc| } 
    \hline
    Module Name& Block name &Configurations\\
    \hline
    \multirow{2}{4em}{Encoder} & Conv1d\_1  &(15,64)$\times 1$\\ 
    & Conv1d\_2 &(5,D)$\times 1$\\ 
    \multirow{1}{4em}{Decoder} & Conv1d\_3  &(41,C)$\times 1$\\
    \multirow{3}{4em}{Phoneme Recognizer} & Conv1d\_4  &(5,64)$\times 3$\\
    & Bi-LSTM & (256)$\times3$\\
    & Linear & (128)$\times2$\\
    \hline
    \end{tabular}
    \end{threeparttable}
    \end{adjustbox}
    \caption{Model Details.   For Conv1d Block, the configuration is (kernel size, output channels)$\times$\# of layers.  For Bi-LSTM and Linear layers, the configuration is (output dimension)$\times$\# of layers.the number of gestures $D$ is a hyperparameter and $C$ is 12. For all convolutional layers, the stride is 1 and paddings are made so that the output length keeps the same across layers. Batchnorm1D \cite{batchnorm} is applied after each convolutional layer.}
    \label{model_config}
\end{table}
\vspace{-10pt}

\subsection{Implementation Details}
For EMA resynthesis experiments, we randomly extract a segment with fixed length of 300 frames as the input of model for each iteration. For phoneme recognition experiments, the full utterance is taken as input. All experiments were trained on Nvidia Tesla V100 GPU. It takes one GPU hour to run a single EMA resynthesis experiment and 5GPU hours to run phoneme recognition as well as resynthesis-CTC experiments. Optimizer is Adam \cite{adam} with the initial learning rate of 1e-3, which is decayed every 5 epoches with a factor of 5. Weight decay is 1e-4. Batchsize is 8. The weights of decoder (gestures) are initialized by the centers from a kmeans algorithm: Slide the window of size 41 with a stride of 1 on EMA kinematics data, concatenate all 41 vectors into a supervector and perform kmeans on all supervectors. For the loss function in Eq. \ref{loss resynthesis} and Eq. \ref{loss joint}, we set $\lambda_1=\lambda_2=\lambda_3=10$ and $\lambda_3=1$. For resynthesis and resynthesis-CTC experiments, we explore different values of number of gestures: 20, 40, 60 and 80 as ablation studies. The results of EMA resynthesis, joint resynthesis-CTC and independent PR on EMA and melspectrogram are recorded in Table. \ref{Tab: Resynthesis-CTC} and Table. \ref{Tab: Independent PR} respectively. To interpret the gestures and gestural scores, a set of utterances are fed into the encoder-decoder framework and we visualize the gestural scores as well as activated gestures. We perform subjective analysis for each utterance and also observe the consistency of the phonetic units across different utterances. A tiny example which takes "Five" as input is  shown in Fig. \ref{model}.
\begin{table}[!ht]
    \centering
    \caption{Resynthesis and Resynthesis-CTC}
    \begin{adjustbox}{width=150pt,center}
    \begin{tabular}{||c c c c c||} 
     \hline
     \hline
     $\#$gestures & 20 & 40 & 60 &80\\ [0.5ex] 
     \hline
     \hline
     \multicolumn{5}{||c||}{Resynthesis}\\
     \hline\hline
     Rec Loss $L_{rec}$ $\%$ & 27.16 & 25.17 & 24.17& 22.99\\ 
     Sparsity $S_1(H)$ $\%$ & 94.10 & 94.50 & 94.17 & 94.90\\
     Sparsity $S_2(H)$ $\%$ & 92.09 & 91.78 & 93.77 & 90.66\\
     \hline\hline
     \multicolumn{5}{||c||}{Resynthesis-CTC}\\
     \hline
     Rec Loss $L_{rec}$ $\%$ & 24.70 & 19.65 & 18.95& 17.72\\ 
     Sparsity $S_1(H)$ $\%$ & 92.90 & 92.54 & 93.10 & 92.50\\
     Sparsity $S_2(H)$ $\%$ & 90.01 & 90.52 & 92.99 & 91.33\\
    PER $\%$ & 20.75& \textbf{14.10} & 15.44& 15.71\\ 
    PER-V $\%$ &16.55  & \textbf{11.02} & 11.88 & 12.09\\
     \hline
    \end{tabular}
    \label{Tab: Resynthesis-CTC}
    \end{adjustbox}
\end{table}

\begin{table}[!ht]
    \centering
    \caption{PER on EMA and Melspec}
    \begin{adjustbox}{width=100pt,center}
    \begin{tabular}{||c c c||} 
     \hline
     Feature & EMA & Melspec \\ [0.5ex] 
     \hline\hline
    PER  $\%$&13.27 & 7.54\\ 
    PER-V $\%$ & 10.24 & 6.18\\
     \hline
    \end{tabular}
    \end{adjustbox}
    \label{Tab: Independent PR}
\end{table}
\vspace{-5pt}

\subsection{Discussion}
We discuss the results in terms of five aspects of the learned gestural scores: Informativeness, Intelligibility, Efficiceny, Interpretability and Consistency.
\paragraph*{Informativeness} Lower reconstruction loss shows that the gestural scores are more informative. By making the comparison between the input EMA and synthesized EMA, we empirically observe that the reconstruction loss that is below 40$\%$ would not loss too much information. As shown in Table. \ref{Tab: Resynthesis-CTC}, the larger the number of gestures, the more informative the gestures are. As it is not hard to overfit the EMA data with an auto-encoder architecture, however, we apply hard sparsity constraint which makes reconstruction much more challenging.

\paragraph*{Intelligibility} Based on Table. \ref{Tab: Independent PR}, EMA gives higher PER and PER-V than melspectrogram because EMA data is sparsely collected from articulators. PER-V of EMA is lower than PER, which is consistent to the fact that EMA is not able to differentiate voiced and voiceless phones. Based on Table. \ref{Tab: Resynthesis-CTC}, when number of gestures is 40, both PER and PER-V are comparable to the results obtained from EMA representation, which shows that gestures scores are intelligible. Note that when increasing the number of gestures, the PER is not always decreasing, indicating that the intelligibility is not always positive correlated to the informativeness. We believe that better intelligibility will be achieved when using more fine-grained articulatory representations such as~\cite{rtMRI}. 

\paragraph*{Efficiency} Based on Table. \ref{Tab: Resynthesis-CTC}, when number of gestures is 40, the sparsity of gestural scores in both time and channel dimension is more than $90\%$, however, the intelligibility does not degrade too much in comparison to the original EMA representation (13.27 versus 14.10 for PER and 10.24 versus 11.02 for PER-V). As mentioned in \textbf{Intelligibility}, if more fine-grained articulatory representations such as~\cite{rtMRI} are applied, it would be possible to still achieve the gestural scores that as intelligible as melspectrogram but much more efficient. 

\begin{figure}[htp]
    \centering
    \includegraphics[width=8.4cm]{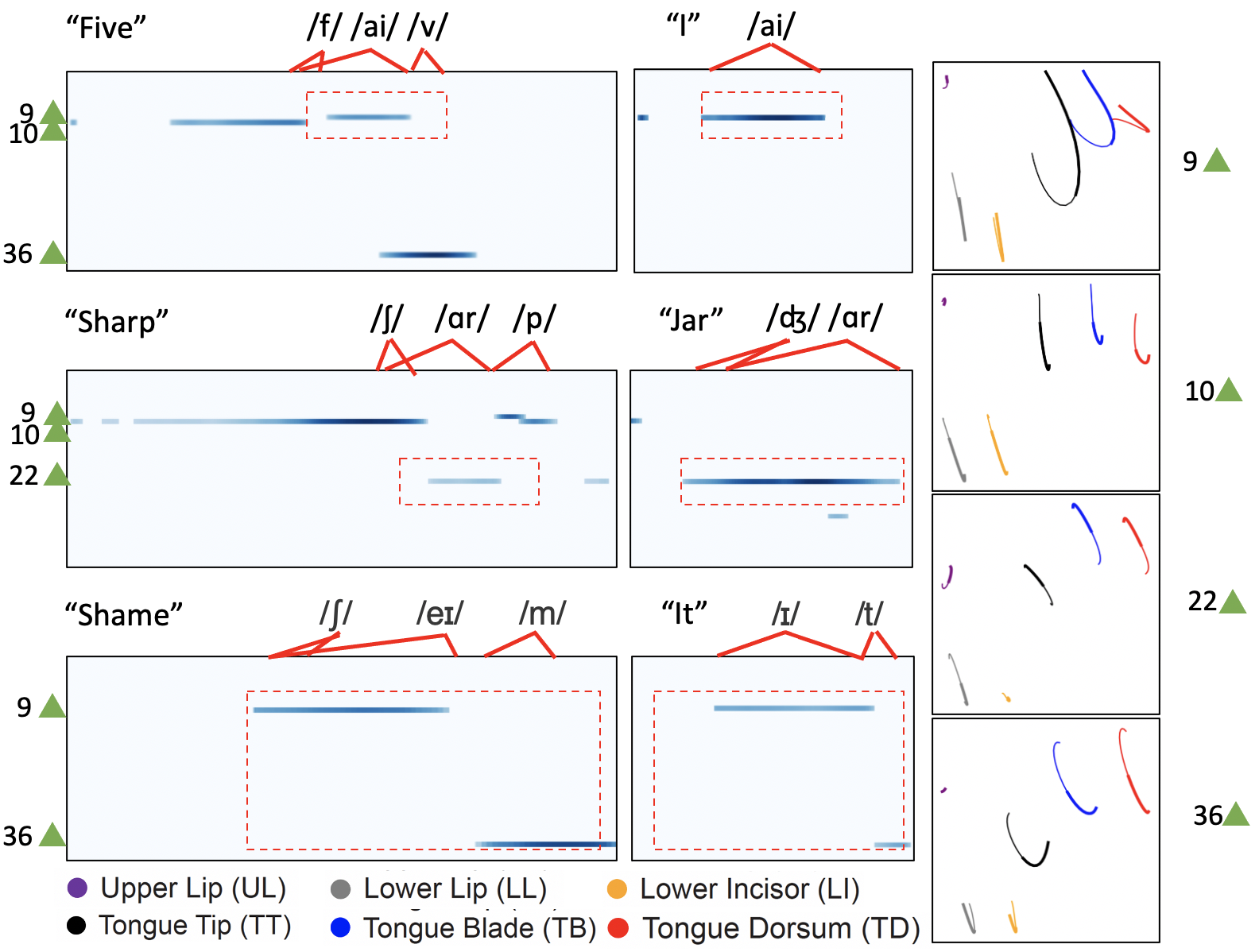}
    \caption{Gestural Scores Visualization.}
    \label{fig:results}
\end{figure}

\paragraph*{Interpretability and Consistency} 
We pick the resynthesis-CTC model with the best phoneme recognition performance ($\#$gestures is 40). Three pairs of short utterances(words) are passed to generate the gestural scores. We visualize both gestural scores and gestures that are activated during the pronunciation. In each gesture, the moving pattern of articulators goes from thinner to thicker. The results are presented in Fig.~\ref{fig:results}. The left six figures are the gestural scores and four gestures that are activated. The solid red line denotes the rough estimation of phoneme duration. The first pair is "Five" and "I" which share the same vowel /ai/. Looking at the word "I", only gesture 9 is activated and gesture 9 exactly carries the moving patterns of "I": when pronouncing "I", we first lower the lower lip, lower incisor and tongue articulators and then raise them. When looking at gestural score of "Five", we observe consistent pattern for /ai/, which is still activated at gesture 9, as indicated by the red block. Other than /ai/, the phoneme /f/ is activated at gesture 10, where all articulators except for upper lib are moving down, which reflects the real articulatory patterns of /f/. For the phoneme /v/, gesture 36 is mainly activated. In gesture 36, all articulators except for upper lip and tongue dorsum first move down and then move up, which exactly reflects the articulatory patterns of /v/. When looking at "sharp" and "jar",  it is observed that /ar/ is  the consistent pattern that is activated at gesture 22, where both the lower lip and lower incisor move down and tongue articulators move up,  which exactly reflects the articulatory patterns of /ar/. For /$\int$/ and /p/, they are mainly activated at gesture 10 where each articulator is lowered. When looking at "shame" and "it", we also observe consistent patterns for /ei/ and /i/, which are mainly activated at gesture 9. The phoneme /$\int$/ is also activated at gesture 9. Note that in gesture 9, each articulator first moves down and then moves up. Such pattern is able to model different phoneme units depending on how it is activated. If the latter part (thicker) of the gesture 9 is activated, it indicates that both the lower lip and lower incisor move down and the tongue articulators move up, which corresponds to the articulatory patterns of /$\int$/. If the former part (thinner) of gesture 9 is activated, it indicates that we just lower all articulators, which corresponds to /ei/ and /i/. Note the articulatory patterns for /m/ and /t/ are also consistent. When gesture 36 is activated, we first lower all articulators except for upper lip and then raise them. In conclusion, both gestures and gestural scores are interpretable and the gestural scores also leverage consistent articulatory pattern across utterances. 
\vspace{-5pt}
\vspace{-13pt}

\section{Conclusion and Limitations}
This work proposes a neural convolutive sparse matrix algorithm which decomposes the EMA data into gestures and gestural scores. The learned representations a.k.a gestural scores are informative, intelligent, consistent, efficient and interpretable. This method bridges the gap between articulatory phonology and deep learning techniques. Hopefully the proposed work could become a paradigm that benefits the downstream explorations that are helpful for patients with vocal cord disorders. One limitation is that EMA data is sparsely sampled from articulators and the learned representation is still less intelligible than the acoustic features. The future work will focus on fine-grained articulatory representations such as~\cite{rtMRI} to deliver more generalizable and intelligible representations. 
\clearpage

\bibliographystyle{IEEEtran}

\bibliography{mybib}

\end{document}